\documentclass[preprint, amssymb, amsmath, nobibnotes, nofootinbib, aps, prb,showpacs]{revtex4-1}
\usepackage[dvips]{graphicx}
\usepackage{epstopdf}
\begin{document}
\title{The critical role of stereochemically active lone pair in introducing high temperature ferroelectricity}
\author{Rafikul Ali Saha$^1$}
\author{Anita Halder$^2$}
\author{Desheng Fu$^3$}
\author{Mitsuru Itoh$^4$}
\author{Tanusri Saha-Dasgupta$^{2}$}
\author{Sugata Ray$^{1}$}
\email{mssr@iacs.res.in}
\affiliation{$^1$School of Materials Sciences, Indian Association for the Cultivation of Science, 2A \& 2B Raja S. C. Mullick Road, Jadavpur, Kolkata 700032, India}
\affiliation{$^2$Department of Condensed Matter Physics and Material Sciences, S. N. Bose National Centre for Basic Sciences, Block JD, Sector 3, Saltlake, Kolkata 700106, India}
\affiliation{$^3$Department of Electronics and Materials Science, and Department of Optoelectronics and Nanostructure Science, Graduate School of Science and Technology, Shizuoka University, 3-5-1 Johoku, Naka-ku, Hamamatsu 432-8561, Japan}
\affiliation{$^4$Materials and Structures Laboratory, Tokyo Institute of Technology, 4259 Nagatsuta, Yokohama 226-8503, Japan}
\pacs {}
\begin{abstract}
In this paper a comparative structural, dielectric and magnetic study of two langasite compounds Ba$_3$TeCo$_3$P$_2$O$_{14}$ (absence of lone pair) and Pb$_3$TeCo$_3$P$_2$O$_{14}$ (Pb$^{2+}$ 6$s^2$ lone pair) have been carried out to precisely explore the development of room temperature spontaneous polarization in presence of stereochemically active lone pair. In case of Pb$_3$TeCo$_3$P$_2$O$_{14}$, mixing of both Pb 6$s$ with Pb 6$p$ and O 2$p$ help the lone pair to be stereochemically active. This stereochemically active lone pair brings a large structural distortion within the unit cell and creates a polar geometry, while Ba$_3$TeCo$_3$P$_2$O$_{14}$ compound remains in a nonpolar structure due to the absence of any such effect. Consequently, polarization measurement under varying electric field confirms room temperature ferroelectricity for Pb$_3$TeCo$_3$P$_2$O$_{14}$, which was not the case of Ba$_3$TeCo$_3$P$_2$O$_{14}$. Detailed study was carried out to understand the microscopic mechanism of ferroelectricity which revealed the exciting underlying activity of poler TeO$_6$ octahedral unit as well as Pb-hexagon.
\end{abstract}
\maketitle
\section{Introduction}
The stereochemically active cationic lone pair driven ferroelectricity within a polar unit cell has always been an important point of investigation in materials physics~\cite{Walsh}. On top of it, the room temperature ferroelectricity is even more fascinating in terms of both fundamental understanding and modern technological applications. In addition, when the systems also accommodate magnetic cations, possibility of multiferroicity arises. In quite a few systems this mechanism seems to work, e.g., ferroelectric perovskites (PbTiO$_3$, BiMnO$_3$, BiFeO$_3$, SnTiO$_3$ etc)~\cite{Jin, Hill, Seshadri, Volkova, Nakhmanson}, double perovskites (Pb$_2$ScTi$_{0.5}$Te$_{0.5}$)O$_6$, Pb$_2$ScSc$_{0.33}$Te$_{0.66}$O$_6$~\cite{Alonso}, Pb$_2$MnWO$_6$~\cite{Mathieu}), as well as $\alpha$-PbO ~\cite{Watson}, SnO, BiOF~\cite{Ramseshadri}, Bi$_2$WO$_6$~\cite{Stolen}, CsPbF$_3$~\cite{Fennie}, BiMn$_2$O$_5$~\cite{Shukla} etc.
\par
Surprisingly, langasite family of compounds have been ignored in this context until recently~\cite{Rafikul}. Thus in order to explore the role of lone pair in bringing ferroelectricity within langasite systems, here in this work we have dealt with two such compounds: Pb$_3$TeCo$_3$P$_2$O$_{14}$ (PTCPO), with lone pair and Ba$_3$TeCo$_3$P$_2$O$_{14}$ (BTCPO), without lone pair. Then we move on to establish that only the presence of Pb$^{2+}$ 6$s^2$ lone pair is not sufficient to bring polarization in the system, the stereochemical activity of the lone pair is the driving force to induce ferroelectricity in the Pb compound. The stronger Pb-O covalent interaction, considerable hybridization of Pb 6$s$ with 6$p$ and Pb 6$s$ with O 2$p$, eventually facilitate the stereochemical activity of the lone pair.
\par
In this paper, we have discussed the structural difference between BTCPO and PTCPO, while BTCPO is noncentrosymmetric and nonpolar, PTCPO is polar. The stereochemically active lone pairs of Pb$^{2+}$ distort the general non polar structure into a polar one. Further, the Te-O covalency creates dissimilar Te-O bond lengths within TeO$_6$ octahedra of PTCPO, helping to enhance the polarization in the Pb containing compound.  In-depth magnetization study of both the compounds clearly reveals the presence of antiferromagnetic transition at low temperature. Further polarization studies under varying electric field at room temperature confirm development of finite polarity in PTCPO. Additionally, the stereochemical activity of the Pb lone pair for the PTCPO compound has been corroborated through spin polarized density functional theory (DFT), crystal orbital Hamiltonian population (COHP) and electron localization function calculations.
\section{Methodology}
\subsection{Experimental techniques}
Polycrystalline samples of Ba$_3$TeCo$_3$P$_2$O$_{14}$ (BTCPO) and Pb$_3$TeCo$_3$P$_2$O$_{14}$ (PTCPO) have been prepared by using conventional solid state reaction techniques. BTCPO was synthesized by taking stoichiometric amounts of high purity BaCO$_3$ (Sigma-Aldrich 99.999\%), TeO$_2$ (Sigma-Aldrich 99.9995\%), Co$_3$O$_4$ (Sigma-Aldrich 99.99\%) and NH$_4$H$_2$PO$_4$ (Sigma-Aldrich 99.999\%), while starting materials for PTCPO were PbO (Sigma-Aldrich 99.999\%), TeO$_2$, Co$_3$O$_4$ and NH$_4$H$_2$PO$_4$. In case of BTCPO, the mixtures were calcined at 600$^{\circ}$~C for 12 hours and finally sintered at 800$^{\circ}$~C for 96 hours under flowing oxygen, while PTCPO compound was prepared by heating the mixture of the starting materials at 600$^{\circ}$~C for 12 hour followed by 700$^{\circ}$~C for 48 hours in oxygen atmosphere in a covered alumina crucible in oxygen atmosphere.
\par
Room temperature structural characterizations for both compounds were carried out by using synchrotron x-ray radiation facility of MCX Beam-line at the Elettra Scnchrotron Centre, Italy with x-ray wave length of 0.827 \AA. Temperature dependent x-ray diffraction (XRD) were carried out at RIGAKU Smartlab (9KW) XG equipped with Cu $K_{\alpha}$ to probe the presence of temperatutre dependent structural phase transitions. The x-ray diffraction data were analyzed via Rietveld refinement using FULLPROF program~\cite{Carvajal}. X-ray photoelectron spectroscopy (XPS) experiments were done in an Omicron electron spectrometer, equipped with SCIENTA OMICRON SPHERA analyzer, Al $K_{\alpha}$ monochromatic source (Model no: XM 1000) and 7 channel channeltron detector. The $dc$ magnetization measurements in the temperature range 2-300 K and in magnetic fields upto $\pm$ 5 Tesla were carried out using a superconducting quantum interference device (SQUID) magnetometer (Quantum Design, USA). The dielectric permitivities of both compounds were performed using a Hewlett-PacPrecision LCR meter (HP4284A) at an $ac$ level 1 V mm$^{-1}$. A cryogenic temperature system (Niki Glass LTS-250-TL-4W) was used to control the temperature within the range of 4 - 450 K. The dielectric hysteresis loops were measured using a ferroelectric measurement system (Toyo Corporation FCE-3) equipped with an Iwatsu ST-3541 capacitive displacement meter having a linearity of 0.1\% and a resolution of 0.3 nm.

\subsection{Theoretical Techniques}
All the first principles density function theory (DFT) calculations have been performed using plane wave pseudopotential method as implemented in Vienna Ab initio Simulation Package (VASP)~\cite{VASP}. The exchange correlation functional has been considered within generalized gradient approximation (GGA)\cite{gga} and to capture the strong electronic correlation at transition metal sites beyond GGA, additional GGA+$U$ calculations have been carried out as prescribed in  Liechtenstein formulation~\cite{Lie}. The $U$ = 5 eV is considered at Co site following the commonly used $U$ value for late transition metals \cite{npj} and the value of Hund's coupling $J_H$ is 0.8 eV. The results have been cross checked by varying the $U$ value within 1-2 eV. The projected augmented wave(PAW) potential~\cite{PAW} has been used and the kinetic energy cut off, set at 600 eV, has been found to provide a good convergence of total energy. Reciprocal-space integration has been carried out with a $k$-space mesh of 2$\times$ 2$\times$ 8. To obtain the energy resolved visualization of chemical bonding, the Crystal Orbital Hamilton Population (COHP)~\cite{COHP} scheme has been considered. A custom software~\cite{lobster} has been employed for this purpose which processes the plane wave calculations by taking the projection into an auxiliary linear combination of atomic orbitals (LCAO) basis and can calculate projected COHP. The tetrahedron method has been used to perform integration in reciprocal space to obtain an energy-resolved COHP plot. To visualize the lone pair arising from the 6$s$ orbital of Pb, we have calculated electron localization functions~\cite{elf1, elf2}. The ELF distribution $\zeta(\overrightarrow{r})$ can be expressed as,

$\zeta(\overrightarrow{r}) = 1/[1 + \chi ^2(\overrightarrow{r})]$ where

$$\chi(\overrightarrow{r}) = \frac{\frac{1}{2}\sum_i \vert \overrightarrow{\nabla}\psi_i\vert^2 - \frac{1}{8}\frac{(\overrightarrow{\nabla}\rho)^2}{\rho}}{\frac{3}{10}(3\pi^2)^{2/3}\rho^{5/3}}$$

where $\rho$ is the electron density and $\psi_i$ is Kohn-Sham wavefunctions. ELF measures the probability of finiding an electron in the vicinity of another electron with same spin and it is used to visualize the atomic shells. The ELF can have value between 0 and 1, where perfect localization is obtained for ELF $=$ 1.

\section {Results and Discussions:}
Rietveld refined room temperature X-ray diffraction (XRD) patterns of both BTCPO and PTCPO compounds are shown in Fig. 1(a) and (b) respectively. BTCPO is fitted by considering a noncentrosymmetric and nonpolar space group $P$321, while PTCPO has been fitted with polar $P$2 space group, which are consistent with previous literature reports~\cite{Krizan, Silverstein}. Refined lattice parameters and goodness factor for both compounds are listed in the table I.  The atomic positions of BTCPO and PTCPO are given in the Table-S1 and S2 in the supplementary material respectively~\cite{Supplementary}. Unit cells of both the crystal structures are shown in the Fig. 1(c) and (d) respectively. Although, like general langasite, Te forms octahedral network, Co and P are in tetrahedral coordination while Pb/Ba are in decahedral arrangement, Pb brings large structural distortion in PTCPO (relative to BTCPO one) by accommodating Pb 6$s^2$ lone pair within the structure. This indeed causes following structural differences between the two compounds:

1. BTCPO takes the simple trigonal $P$321 structure with the Ba-Te layers defining the unit cell boundaries (see the $a-c$ plane of the unit cell in Fig. 2(a)). For PTCPO, the Pb lone pair brings forth distortion in the unit cell where few of the Co and Pb ions move away slightly along the $c$ axis, thereby reducing the symmetry and a polar, monoclinic $P$2 space group is realized.

2. BTCPO possesses single equilateral Co and Ba triangles in each unit cell, while PTCPO intakes isosceles triangular networks formed by Co and Pb, as shown in Figs. 2(c) and (d).

3. The equilateral triangles of BTCPO remain perfectly in the $a-b$ plane, but in case of PTCPO the isosceles triangles of Pb and Co undergo a relative tilting with respect to each other, as indicated in Figs. 2(e) and (f). The tilting of the triangles away from the $a-b$ plane (up or down of the plane) depends on the movement of Pb and Co atoms along $c$-axis ('+' or '-' ve $c$-direction).

4. Unlike BTCPO, the centre of mass of anionic network (0, 0.334, 1) within the TeO$_6$ octahedra does not coincide with the position of Te (0, 0.336, 1) in PTCPO.

As a result of these subtle structural differences, PTCPO takes much bigger unit cell (six times) compared to BTCPO. However, it is primarily important to confirm the existence of stereochemically active Pb$^{2+}$ lone pair at room temperature. In the context of Pb lone pair, the degree of stereochemical activity of lone pair is measured by calculating the values of $\triangle$$E_1$ (Difference between the shortest distances Pb-O from the spheres I and II) and $\triangle$$E_2$ (difference between the shortest distance Pb-O in the polyhedron of the compound under study and the known shortest distance in compounds containing these elements)~\cite{Volkova}. $\triangle$$E_1$ = 0.32 {\AA} and $\triangle$$E_2$ = 0.05 {\AA} (considering the minimal distance $d$(Pb$^{2+}$-O) = 2.40 {\AA} (as in Pb$_2$V$_5$O$_{12}$)~\cite{Volkova} specify the high degree of stereochemical activity of Pb$^{2+}$ lone pair in PTCPO. In addition, nonuniform Te-O bond distances within TeO$_6$ octahedra of PTCPO suggest presence of charge sharing variations between Te and O, further helping to get a polar structure.

\par
Further x-ray photoelectron spectroscopy (XPS) technique is used to probe the proper oxidation states of the elements of both the compounds. We probed the core level Pb 4$f$, Te 3$d$, V 2$p$, P 2$s$, Co 2$p$ and O 1$s$ for both samples and the corresponding fitting have been performed. The energy positions of the respective features clearly indicate the expected charge states of all the cations for both these samples (not shown here). While O 1$s$ core level spectra of BTCPO shows a clear singlet, PTCPO possesses asymmetric broadening in the higher binding energy side of the O 1$s$ spectrum, as shown in Figs. 3(a) and (b). This spectrum has been fitted by considering two singlets, which signifies strong variation in the electron density of the oxygens of this compound, which results in two distinctly different O 1$s$ signals, compared to BTCPO~\cite{Bandyopadhyay}.

\par
We have carried out density functional theory (DFT) calculations in order to assess the covalency effect in the context of stereochemical activity of the Pb lone pair. Since the goal of our first-principles calculation is to reveal the covalency effect between different cations and oxygen, it is realized that the analysis of electronic structure should be carried out within the realm of spin-polarized ferromagnetic calculations. The antiferromagnetic structure, which is the ground state magnetic structure is complex and leads to cancellation of covalency effect due to opposite alignment of spins, as ascertained in vanishing moment of oxygen compared to its finite value in fully spin-polarized calculation. The calculations have been carried out both for the BTCPO (nonpolar $P$321) and PTCPO (polar $P$2) structures and the detailed density of states (DOS) are shown in Figs. 4(a) and (b), respectively. DFT calculation shows presence of Ba $s$ partial density of states (PDOS) above the Fermi energy and absence of PDOS below the Fermi energy signifying the presence of pure Ba$^{2+}$ in the system. In case of PTCPO, Pb 6$p$ and 6$s$ PDOS are seen above (3 to 8 eV) and below (down to -9 eV) the Fermi level respectively, supporting the presence of Pb$^{2+}$ state. Interestingly, presence of both Pb 6$s$ and 6$p$ PDOS in the energy range -6 to -1 eV  indicates the Pb 6$s$-6$p$ mixing in the compound. The stereochemical activity of lone pair active ions like Bi$^{3+}$ or Pb$^{2+}$ within oxide structures has been argued to originate from the hybridization between 6$s$, 6$p$ orbitals at Bi/Pb site and 2$p$ orbitals at O site~\cite{Atanu, Watson2}. The 6$p$-2$p$ hybridization turns out to be the governing factor in giving rise to directionality of the lone pair and driving the off-centric distortion of the unit cell~\cite{Walsh, Rafikul}. Here, strong hybridization between Pb 6$s$, 6$p$ and O 2$p$ gets confirmed, as indicated in the cyan shaded region. Additionally, presence of finite DOS from Pb 6$s$, Te 6$p$ and O 2$p$ in the energy range of -6.5 to -8 eV signifies the hybridization of Pb and Te with same O. But in case of BTCPO, magenta shaded region signifies relatively weaker hybridization of Te and Co with the oxygen. Further, this has been quantitatively confirmed from our COHP (Crystal Orbital Hamilton Population) calculation where the integrated COHP (ICOHP) values have been found to become larger for the cation-oxygen connectivity while going from BTCPO ($P$321) to PTCPO ($P$2) (see Figs. 4(c)-(d) and Table II). In this context, the elemental magnetic moment presented a noticeable change in the PTCPO structure relative to the BTCPO, in line with the stronger covalency effect in PTCPO (see Table II). Such an effect should obviously enhance the degree of Pb lone pair activity in the compound at lower temperature. The charge density distributions of BTCPO confirm spherical symmetric distribution of Ba, while lobe like structures are observed for PTCPO, as shown in Figs. 4(e) and (f). This lobe like structure of Pb 6$s^2$ lone pair appears due to the mixing of Pb 6$s$ with 6$p$ and Pb 6$s$ with O $p$, signifying the stereochemical activity of Pb lone pair~\cite{Atanu, Watson2}. In addition, asymmetric Te lobes might indicate non uniform Te-O covalent interactions, and consequently non uniform Te-O bond distances. As a result, we may conclude that the covalency effect of both Pb-O and Te-O may give rise to spontaneous polarization in the system. Further, hybridization of Pb and Co with same O and also Pb and Te with same O firmly suggest the stereochemical active Pb lone pair to be responsible for the movement of Co triangles and TeO$_6$ octahedra in this sample, as indicated in Fig. 2(a)-(f).


\par
The temperature dependence of the real part of the dielectric constant ($\varepsilon'$/$\varepsilon_0$) and dielectric loss ($\tan\delta$) have been carried out in the temperature range of 5 to 350 K for BTCPO and 5 to 400 K for PTCPO at different frequencies (20 kHz-1000 kHz), as shown in Fig. 5(a) and (b) respectively. In case of PTCPO, a frequency independent anomaly  near 365 K appears in the temperature dependent dielectric constant and loss ($\tan\delta$) data, which is further supported by the respective first order derivative curves (see Fig. S1 of the supplementary materials~\cite{Supplementary}). On the other hand, there is no anomaly in the measured temperature range of BTCPO. In order to identify the origin of dielectric anomaly, temperature dependent X-ray diffraction measurements of PTCPO compound have been performed over a wide temperature range of 4-400 K. The collected XRD patterns have been fitted using monoclinic space group $P$2. An anomaly near 365 K is observed in the temperature dependent lattice parameters as well as unit cell volume variations, which coincide with the anomaly of temperature dependent dielectric variation (see Fig. 5(c)) signifying the development of ferroelectricity below that temperature. The expanded view of high temperature (above 365 K) and low temperature XRD patterns are shown in the Fig. S2 of the supplementary material~\cite{Supplementary} which reveals development of many small peaks below $T_C$. However, due to only minor changes in position coordinates across the phase transition, the present Rietveld refinements could not conclusively determine the space group of high temperature phase. This is not very uncommon in presence of weak lattice distortion~\cite{Sampathkumaran1, Spaldin3}. Further polarization ($P$) with varying electric field ($E$) at room temperature (300 K) has been measured for both BTCPO and PTCPO compounds, as highlighted in Fig. S3 of the supplementary material~\cite{Supplementary} and Fig. 6(a) respectively. Clear $P-E$ loop with remnant polarization of 0.35 $\mu$C/cm$^2$ and a butterfly loop in strain versus electric field curve are observed in PTCPO which support the true ferroelectric ordering at room temperature. It is to be noted that PTCPO is a good insulator because the sample can bear a high field of 130 kV/cm, which ensures no leakage. The microscopic origin of the electrical dipoles in the PTCPO system can be clearly understood in terms of the strength ($\bigtriangleup E_1$) of stereochemically active Pb lone pair and local polar geometry of TeO$_6$ octahedra. The value of $\bigtriangleup E_1$ becomes 0.17 at 400 K, which is significantly much smaller than the same at the 300 K (0.32). This implies weakening of the stereochemical activity of Pb lone pair above ferroelectric $T_C$. Moreover, at 300 K the center of mass of Te cation (0, 0.336, 1) moves away from the anionic center of mass (0, 0.334, 1) along the $y$-coordinates, this difference in center of masses between Te (0, 0.3279, 1) and the anions (0, 0.3277, 1) becomes negligible at 400 K. The development of spontaneous room temperature polarization in PTCPO could be attributed firstly to the stereochemically active lone pair of Pb$^{2+}$ ($ns^2$) in Pb-Pb hexagon and secondly to the polar geometry of TeO$_6$ octahedra. Therefore, a resultant dipole moment gets developed within a single Pb$_6$ motif (where Te and Pb are present in the $a-b$ plane but tilted along $c$ direction) along the $b-c$ plane, as shown in Fig. 6(b). However, the small measured spontaneous polarization (~0.5 $\mu$C/cm$^2$), structural data and the type of dielectric anomaly as well as multiplication of the unit cell might indicate that the ferroelectric transition in PTCPO is improper. On the other hand, in case of BTCPO, the weaker Co-O covalency within CoO$_4$ tetrahedra, and the consequent reduced charge distribution among the other cation-oxygen bonds, together oppose the geometric polar structure in BTCPO. This eventually negates the room temperature ferroelectricity in this compound, unlike in PTCPO.
\par
Next we have investigated the temperature dependent $dc$ magnetization of BTCPO and PTCPO compounds. The zero field cool (ZFC) and field cool (FC) 500 Oe magnetic susceptibility data of both BTCPO and PTCPO in the temperature range 2-300 K, are shown in Fig. 7(a) and (b) respectively. Both order antiferromagnetically, with Neel temperature ($T_N$) of around 21 K and 14 K for BTCPO and PTCPO respectively, consistent with previous study~\cite{Silverstein, Krizan}. However, no perceptible dielectric anomaly at the antiferromagnetic transition indicates a weak magnetoelectric coupling in PTCPO. Curie-Weiss (CW) fit ( $\chi$ = $\chi_0$ + $C$/($T$ - $\theta_{\text{CW}}$)); where $\chi_0$ is the temperature independent paramagnetic succeptibility, $C$ is Curie constant related to the effective moment, and $\theta_{\text{CW}}$ being Weiss temperature) on the 500 Oe field cooled susceptibility data in the temperature range $T$= 100 - 300 K have been performed for both the samples and the subsequent results are, shown in the inset of Fig. 7(a) and (b). The negative Weiss temperatures ($\theta_{\text{CW}}$ = -25.12 K and -23.1 K for BTCPO and PTCPO respectively), consistent with previous studies~\cite{Krizan}, indicate the presence of antiferromagnetic interactions between Co spins, possibly due to dominant Co-O-O-Co super-superexchange pathways within neighboring Co-triangles~\cite{Krizan}. The effective paramagnetic moments, $\mu_{\text{eff}}$ = 4.74 and 4.43  $\mu_{\text{B}}$/Co for BTCPO and PTCPO respectively, obtained from the C-W analysis, are higher compared to the theoretically calculated spin only Co-moment value (3.87 $\mu_{\text{B}}$), suggesting incomplete quenching of Co-orbital moments.
\section{Conclusion}
In summary, detailed structural, dielectric and magnetic investigations as well as first-principles electronic structure calculations have been performed on two langasite compounds BTCPO and PTCPO. We have confirmed through detailed structural analysis that BTCPO (without lone pair) is non-polar, but PTCPO (lone pair) is polar at room temperature. Stronger Co-O covalency in PTCPO helps to redistribute the charges among the other cation-oxygen bonds, which not only affects the individual elemental magnetic moments but also facilitates the lone pair activity of Pb by enhancing the mixing of Pb 6$s$ with Pb 6$p$ and O 2$p$. On top of it, strong Te-O covalency causes significant charge disproportionation and thus creates nonuniform Te-O bond distances within a TeO$_6$ octahedra. The combinatorial effect of stereochemically active lone pair and Te-O covalency help to induce spontaneous polarization at room temperature in the PTCPO.

\section{Acknowledgement}
R.A.S thanks CSIR, India and IACS for a fellowship. S.R thanks DST, India for funding (project no.CRG/2019/003522), Technical Research Center of IACS, Indo-Italian POC for support to carry out experiments in Elettra, Italy and Laboratory for Materials and Structures, Collaborative Research Projects for providing experimental facilities. A.H and T.S.D acknowledge the computational support of Thematic Unit of Excellence on Computational Materials Science, funded by Nano-mission of Department of Science.

\newpage
\begin{table*}
\caption{Lattice parameter and goodness factor of BTCPO and PTCPO compounds.}
\resizebox{12cm}{!}{
\begin{tabular}{| c | c | c | c | c | c | c | c | c |}\hline
Sample & Space group &  a ({\AA}) & b ({\AA}) & c ({\AA}) &  $\alpha$ & $\beta$ & $\gamma$ & $\chi^2$ \\\hline
BTCPO & $P$321 & 8.470 & 8.470 & 5.316 & 90 & 90 & 120 & 2.81 \\
PTCPO & $P$2 & 14.504 & 25.106 & 5.228 & 90 & 90.02 & 90 & 2.13 \\
\hline
\end{tabular}
}
\end{table*}

\begin{table*}
\caption{ICOHP values and magnetic moments of BTCPO and PTCPO compounds.}
\resizebox{14cm}{!}{
\begin{tabular}{| c | c | c | c | c | c | c | c |}
\hline \multicolumn {4}{c} {Value of ICOHP (eV/bond)} && \multicolumn {3}{c} {Value of magnetic moment in $\mu_B$}\\\hline
Sample & Space group & Ba/Pb-O &  Co-O & Te-O & Co &  Ba/Pb & O \\\hline
BTCPO & $P$321 & -0.069 & 0.338 & 0.272 & 2.733 & 0.002 & 0.034 \\
PTCPO & $P$2 & 0.439 & 0.464 & 0.651 & 2.55 & 0.009 & 0.063 \\
\hline
\end{tabular}
}
\end{table*}

\begin{figure}
\resizebox{8.6cm}{!}
{\includegraphics[73pt,387pt][520pt,770pt]{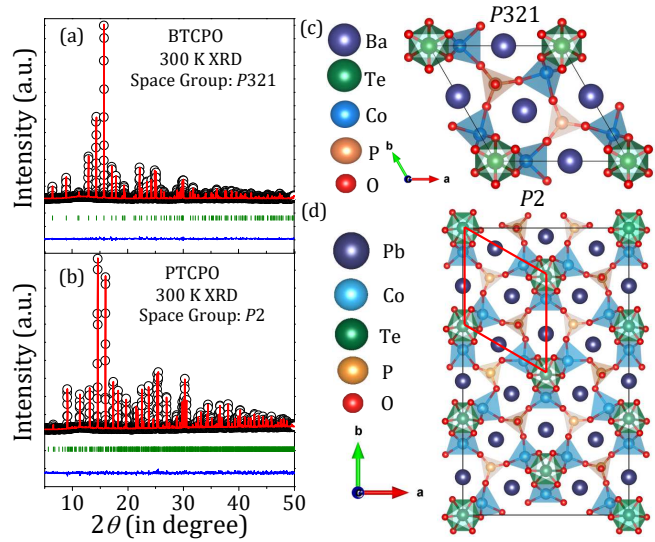}}
\caption{Rietveld refined XRD of (a) BTCPO and (b) PTCPO. Black open circles represent the experimental data and red line represents the calculated pattern. The blue line represents the difference between the observed and calculated pattern and green lines signify the position of Bragg peaks. (c) and (d) are the crystal structure of BTCPO and PTCPO respectively.}
\end{figure}

\begin{figure}
\resizebox{8.6cm}{!}
{\includegraphics[50pt,351pt][522pt,801pt]{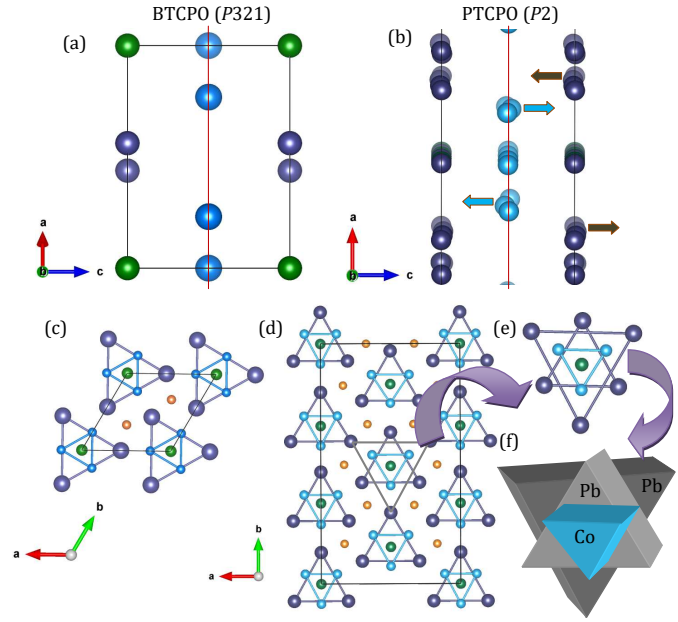}}
\caption{(a) The positions of Pb, Te and Co in BTCPO unit cell. Pb and Te are in the side line of the unit cell, while all Co are exactly in the middle line (red) of the two boundaries. (b) The positions of Pb and Co in PTCPO unit cell. Some Pb are shifted towards positive or negative $c$ axis from  boundary line, represented by brown arrow, and some Co are shifted towards positive or negative $c$ axis from  middle line (red), represented by cyan coloured arrow.    (c) Gray and cyan coloured triangles are the Ba equilateral triangles and Co equilateral triangles of BTCPO respectively. (d) Gray and cyan coloured triangles are the Pb isosceles triangles and Co isosceles triangles of PTCPO respectively. (e) a portion of PTCPO which contains two Pb and one Co isosceles triangles. (f) Schematic representation of two tilted Pb and one tilted Co triangles of PTCPO.}
\end{figure}

\begin{figure}
\resizebox{8.6cm}{!}
{\includegraphics[88pt,607pt][477pt,791pt]{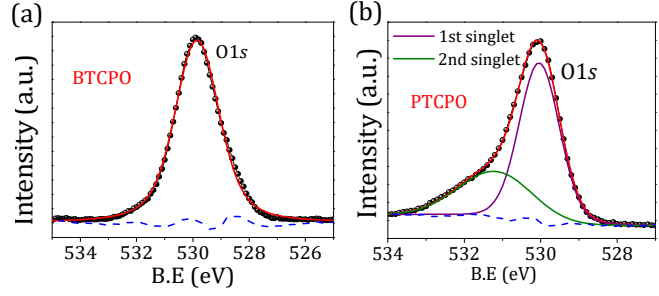}}
\caption{(a) and (b) The Shirley background corrected experimental data (shaded black circles) along with the theoretical fitting (solid red line) and difference between experimental data and theoretically fitted data (blue dotted line) are shown for Oxygen (O 1$s$) of BTCPO and PTCPO respectively.}
\end{figure}

\begin{figure}
\resizebox{8.6cm}{!}
{\includegraphics[40pt,129pt][552pt,783pt]{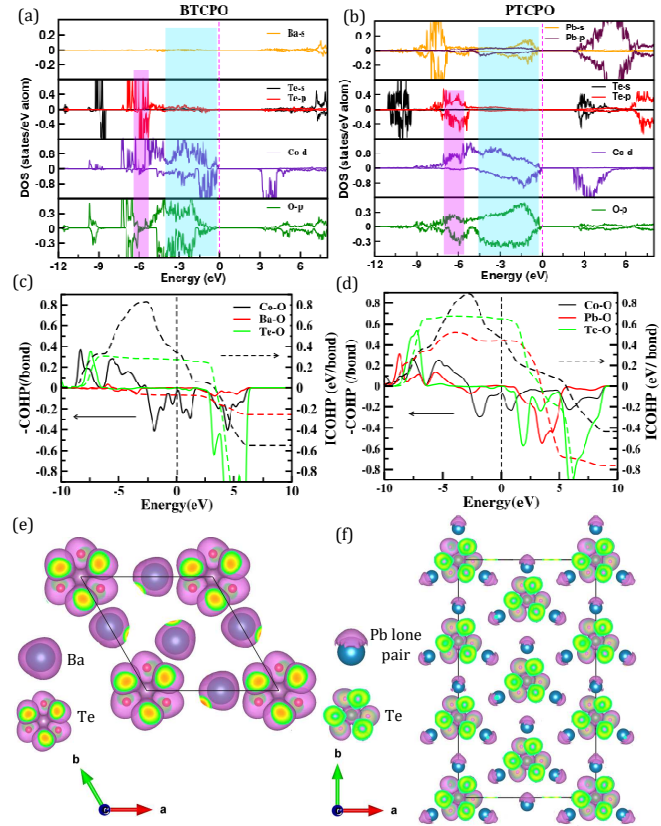}}
\caption{(a) and (b) Partial density of state of BTCPO and PTCPO respectively. (c) and (d) COHP and ICOHP curve of BTCPO and PTCPO respectively. (e) and (f) Electron localization function within a unit cell (The isosurfaces are vizualized for a value of 0.3) of BTCPO and PTCPO respectively.}
\end{figure}

\begin{figure}
\resizebox{8.6cm}{!}
{\includegraphics[107pt,276pt][520pt,781pt]{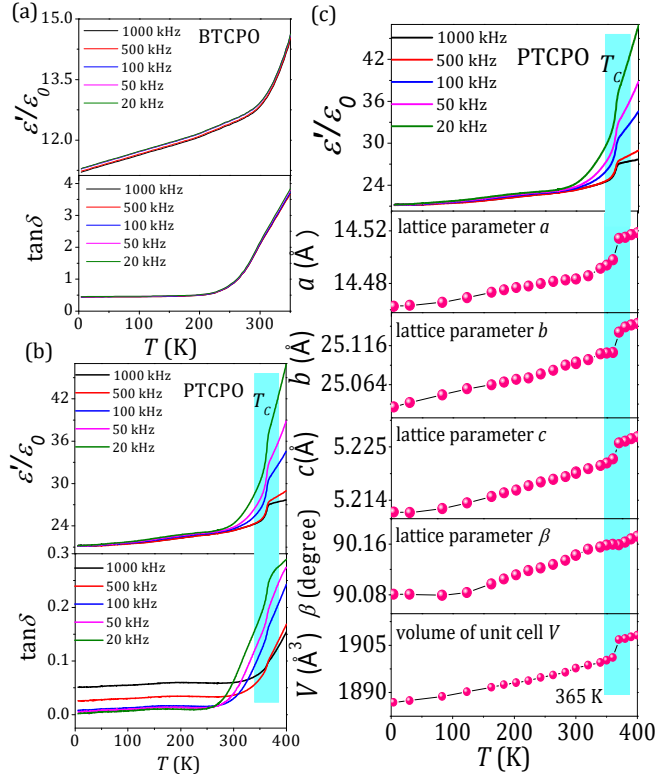}}
\caption{(a) and (b) Temperature dependence of real part of dielectric constant $\varepsilon'$/$\varepsilon_0$ and $\tan\delta$ loss data of BTCPO and PTCPO respectively. (c) Thermal variation of real part of dielectric constant, lattice parameters and volume of unit cell of PTCPO.}
\end{figure}
\begin{figure}
\resizebox{6cm}{!}
{\includegraphics[120pt,353pt][412pt,700pt]{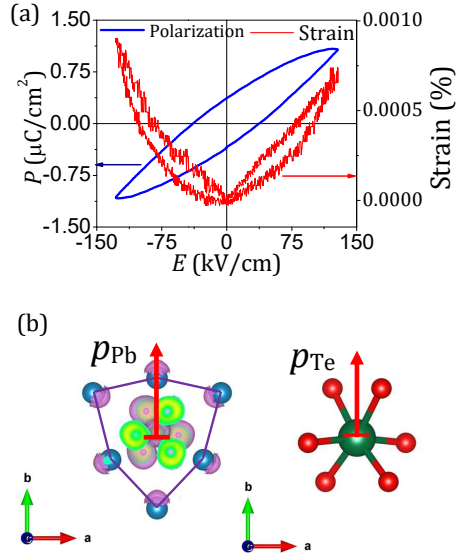}}
\caption{(a) Electric field variation polarization and strain of PTCPO at room temperature. (b) Pb$_6$ hexagon and TeO$_6$ octahedra of PTCPO structure. Pb$_6$ hexagon and TeO$_6$ octahedra give net dipole moment.}
\end{figure}

\begin{figure}
\resizebox{8.6cm}{!}
{\includegraphics[57pt,539pt][485pt,756pt]{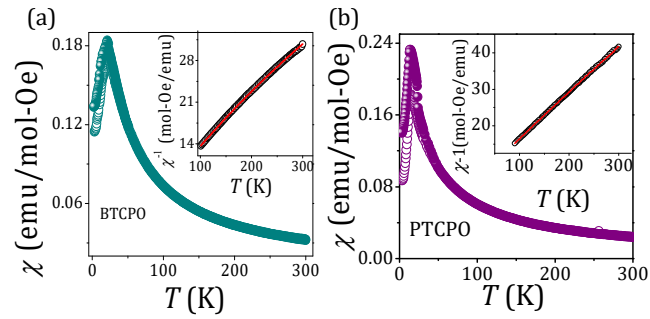}}
\caption{(a) and (b) ZFC and FC at 500 Oe of BTCPO and PTCPO respectively and insets show Curie-Weiss fitting of both compounds. Open circle represents ZFC and solid ball signifies FC curve. Red line is the Curie-Weiss fitted curve.}
\end{figure}
\end{document}